\documentclass[aps,prc,twocolumn,superscriptaddress,showpacs,nofootinbib]{revtex4-1}
\usepackage{graphicx}
\usepackage{amsmath,amssymb}
\usepackage{bm}
\usepackage[colorlinks,citecolor=blue]{hyperref}

\begin{document}

\title{Unblocking of stellar electron capture for neutron-rich $N=50$ nuclei at
finite temperature}

\author{Alan A. Dzhioev}
\email{dzhioev@theor.jinr.ru}
\affiliation{Bogoliubov Laboratory of Theoretical Physics, JINR, 141980, Dubna, Russia}

\author{K. Langanke}
\affiliation{GSI Helmholzzentrum f\"{u}r Schwerionenforschung, Planckstra{\ss}e 1, 64291 Darmstadt, Germany}
\affiliation{Institut f\"{u}r Kernphysik (Theoriezentrum), Technische Universit\"{a}t Darmstadt,
                                       Schlossgartenstra{\ss}e 2, 64289 Darmstadt, Germany}

\author{G. Mart\'{\i}nez-Pinedo}
\affiliation{GSI Helmholzzentrum f\"{u}r Schwerionenforschung, Planckstra{\ss}e 1, 64291 Darmstadt, Germany}
\affiliation{Institut f\"{u}r Kernphysik (Theoriezentrum), Technische Universit\"{a}t Darmstadt,
                                       Schlossgartenstra{\ss}e 2, 64289 Darmstadt, Germany}

\author{A. I. Vdovin}
\affiliation{Bogoliubov Laboratory of Theoretical Physics, JINR, 141980, Dubna, Russia}

\author{Ch.~Stoyanov}
\affiliation{Institute for Nuclear Research and Nuclear Energy, Bulgarian Academy
of Sciences, 1784 Sofia, Bulgaria}

\date{\today}

\begin{abstract}
We have calculated electron capture rates for neutron-rich $N=50$ nuclei
($^{78}$Ni, $^{82}$Ge, $^{86}$Kr, $^{88}$Sr)
within the thermal QRPA approach at temperatures $T=0$, corresponding
to capture on the ground-state, and at $T=10$~GK (0.86~MeV), which is a typical
temperature at which the $N=50$ nuclei are abundant during a supernova collapse.
	In agreement with recent experiments, we find no Gamow-Teller (GT$_+$) strength
at low excitation energies, $E<7$ MeV, caused by Pauli blocking
induced by the $N=50$ shell gap. At the astrophysically relevant temperatures
this Pauli blocking of the GT$_+$ strength is overcome by thermal excitations
across the $Z=40$ proton and $N=50$ neutron shell gaps, leading to a sizable
GT contribution to the electron capture. At the high densities, at which the
$N=50$ nuclei are important for stellar electron capture, forbidden transitions
contribute noticeably to the capture rate. Our results indicate that the
neutron-rich $N=50$ nuclei do not serve as an obstacle of electron capture during
supernova collapse.
\end{abstract}



\maketitle

\section{Introduction}

Electron captures on nuclei play an essential role during the collapse of a massive star leading to a type II or core-collapse
supernova~\cite{Bethe_NPA324,Bethe_RevModPhys62,Langanke_RevModPhys75,Janka_PhysRep442}. It reduces the electron-to-baryon ratio
$Y_e$ and hence the pressure which the relativistic degenerate electron gas can stem against the gravitational collapse. As the neutrinos
produced by the capture process can leave the star, carrying away energy, it is also an effective cooling mechanism, resulting in the fact
that heavy nuclei survive during the collapse~\cite{Bethe_NPA324}. The temperature in the collapsing core is sufficiently
high that nuclei exist in nuclear statistical equilibrium (NSE)~\cite{Hix_ApJ460}.
However, due to the decrease of $Y_e$ by continuous electron captures, the abundance distribution of nuclei is shifted to more neutron-rich
and heavier nuclei during the collapse.

Due to the electron energies involved, electron captures are dominated by allowed Gamow-Teller (GT$_+$) transitions (in which a proton
is changed to a neutron) at the early stage of the collapse. However, forbidden transitions become increasingly important
with growing electron energies and contribute significantly to the capture rates in the later collapse phases~\cite{Cooperstein_NPA420,Juodagalvis_NPA848}.
 For core densities $\rho \lesssim 10^{10}$ g~cm$^{-3}$ and the
respective temperatures the core composition of nuclei is given by $pf$ shell nuclei in the iron-nickel mass region.
For these nuclei electron capture rates can be calculated on the basis of large-scale shell model diagonalization calculations~\cite{Langanke_NPA673,Langanke_ADNDT79,Suzuki_PRC83}.
The calculations reproduce the GT$_+$ distributions experimentally determined by charge-exchange reactions~\cite{Fujita_PPNP66, Frekers_EPJA54}
quite well~\cite{Caurier_NPA653,Caurier_RMPh77,Cole_PRC86}. These rates are significantly smaller
than the pioneering rates by Fuller {\it et al.}~\cite{Fuller_APJS48}, resulting in a slower deleptonization in the early
collapse phase~\cite{Heger_PRL86,Heger_APJ560}.

As noted by Fuller~\cite{Fuller_APJ252} the continuous shift
of the NSE abundance distribution to heavier and more neutron-rich nuclei
can lead to a potential blocking of the GT$_+$ strength, once nuclei
with proton numbers $Z<40$ and neutron numbers $N>40$ dominate the
core composition. For such nuclei, GT$_+$ transitions are completely
Pauli blocked within the simple independent-particle model. Based on this
observation, Bruenn derived stellar capture rates which predicted
vanishing capture rates for nuclei with $N>38$ \cite{Bruenn_ApJS58}. These capture
rates have been the standard in supernova simulations for many years
and led to the conclusion that electron capture proceeds on free protons
in the advanced collapse phases (e.g.~\cite{Bethe_RevModPhys62}). Cooperstein
and Wambach pointed out that the Pauli blocking might be overcome by
thermal excitations, but which would only happen at core densities
in excess of $10^{11}$~g~cm$^{-3}$~\cite{Cooperstein_NPA420}.
However, the $N=40$ shell closure is overcome by cross-shell correlations
which move neutrons and protons into the $g_{9/2}$ orbital and hence open
up GT$_+$ transitions. Experimentally this is observed for $^{76}$Se ($Z=34,~N=42$)
which has a non-vanishing GT$_+$ strength distribution (required
for the double-beta decay of $^{76}$Ge~\cite{Grewe_PRC78,Frekers_EPJA54}), made possible
by a sizable neutron-hole structure in the $pf$ shell as determined from
transfer reactions~\cite{Kay_PRC79}. The experimental GT$_+$ distribution
is well described by shell model diagonalization studies~\cite{Zhi_NPA859}
confirming that cross-shell correlations require multi-particle-multi-hole
correlations \cite{Dean_PRC59,Caurier_PLB522}.
For stellar electron capture rates, these correlations have been considered
within a hybrid model, in which nuclear partial occupation numbers have been
calculated within the Shell Model Monte Carlo Approach~\cite{Johnson_PRL69,
Koonin_PhysRep278} which allows to determine thermally-averaged
nuclear properties at finite temperatures considering
correlations in unprecedentedly large model spaces (here the full $pf-gds$ shells).
The partial occupation numbers served as input to a random-phase approximation (RPA) calculation
of the stellar electron capture rates~\cite{Langanke_PRL90}.
Incorporated into supernova simulations these rates had noticeable effects
on the supernova dynamics \cite{Hix_PRL91,Janka_PhysRep442} and
showed that electron capture is dominated by nuclei during the entire collapse.

Sullivan {\it et al.}~\cite{Sullivan_APJ816,Titus_JPhG45} have pointed out that the $N=50$ shell gap at
the neutron $g_{9/2}$ shell closure could act as a severe obstacle for
stellar electron captures, in particular for nuclei with proton number
$Z<40$ because they are frequently encountered at core densities before neutrino
trapping (at a few $10^{11}$ g~cm$^{-3}$).
The argument is based on the observation
that Pauli unblocking by neutron holes in the $pf$ shell is strongly hindered by the
gap, and that proton excitations into the $g_{9/2}$ orbital would mainly
lead to GT$_+$ transitions into the neutron $g_{7/2}$ orbital residing
at modest excitation energies in the daughter and hence will not noticeably contribute
to the electron capture rate. Sullivan and collaborators supported their argument
by studies with parametrized stellar electron capture rates which
indicated the $N=50$ nuclei as an obstacle to the supernova dynamics
\cite{Sullivan_APJ816,Titus_JPhG45}. Motivated by these studies,
Zegers {\it et al.} measured the GT$_+$ strength distribution
in two relevant nuclei, $^{86}$Kr ($Z=36 , N=50$)~\cite{Titus_arXiv1908} and
$^{88}$Sr ($Z=38, N=50$)~\cite{Zamora_PRC100}. Both distributions indeed show
no GT$_+$ strength at low energies. These authors then used
the experimental GT$_+$ distributions for the nuclear ground states
to determine stellar electron capture rates. Such a procedure would be valid
if the Brink-Axel hypothesis holds, i.e., the GT$_+$ distribution
on all excited states is the same as for the ground state.

As we  show in the following, this assumption is inappropriate.
At first, in the stellar core the capture occurs at finite
temperatures of about $T=1 $ MeV. Adopting the simple Fermi gas ansatz, this
temperature translates into excitation energies ($E^*\approx
AT^2/8 $) of about 10 MeV, which is larger than the shell gaps at $N=50$ and
$Z=40$. Hence, the capture occurs on a thermal nuclear ensemble which includes
excited states with proton particles in the $g_{9/2}$ orbital and neutron holes
in the $pf$ and $g_{9/2}$ orbitals.
These correlations unblock  GT$_+$ transitions at low energies or can even lead
to nuclear deexcitation where nuclear excitation energy is transferred to the leptons.
We note that a sizable unblocking of the GT$_+$ strength by correlations and
thermal excitations was found in the the hybrid SMMC/RPA calculations,
exemplified for $^{89}$Br in Ref.~\cite{Langanke_PRL90}, and
for $^{76-80}$Ge, $^{78}$Ni based on the thermal QRPA (TQRPA) approach~\cite{Dzhioev_PRC81,Dzhioev_PRC100}.
These studies also showed that forbidden transitions, which are not hindered by the
shell gap, contribute sizably to the stellar capture rates at the conditions which are relevant for $N=50$ nuclei.

In this paper we extend the TQRPA study of Dzhioev and collaborators
to a chain of $N=50$ nuclei, including the two nuclei ($^{86}$Kr and $^{88}$Sr)
for which experimental GT$_+$ distributions have been measured for the ground
states. The TQRPA consistently describes thermal properties of nuclei
at finite temperatures considering 2p-2h correlations induced by pairing
and a residual interaction. In the limit of vanishing temperatures,
it reduces to the QRPA model. Our focus is here on the aspect how
the correlations unblock the GT$_+$ strength at finite temperature
and which consequences this unblocking has on the stellar electron capture rate.
We will also calculate the forbidden contributions to the rate. Our main result is
that the capture rate for $N=50$ nuclei at the finite temperatures,
which are relevant in a supernova collapse, is much larger than estimated
on the basis of the GT$_+$ ground state distributions.

We should mention several papers where different finite-temperature RPA models based on  Skyrme and  relativistic energy density functionals  have been used to calculate stellar electron capture (EC) rates~~\cite{Paar_PRC80,Niu_PRC83,Fantina_PRC86}. The TQRPA approach differs from those of Refs.~\cite{Paar_PRC80,Niu_PRC83,Fantina_PRC86} primarily by thermodynamically consistent consideration of thermal effects. It was shown in Ref.~\cite{Dzhioev_PhAN79} that exoergic transitions from thermally excited states appear within the TQRPA and for EC on $^{56}$Fe they remove the reaction threshold and enhance the low-energy cross section. In contrast, no such transitions appear within the finite-temperature RPA models. As a result, calculations in Refs.~\cite{Paar_PRC80,Niu_PRC83,Fantina_PRC86} predict that EC cross sections drop rapidly to zero as the electron energy falls below some threshold value.

Our paper is organized as follows. In the next section we give a brief outline of the TQRPA method
which we have used in our calculations. A comprehensive description of the thermal QRPA approach is given in Refs.~\cite{Dzhioev_PRC81,Dzhioev_PRC100}.
In Section~\ref{results} we discuss the results of our calculations. In Section~\ref{conclusion}  we provide the concluding remarks.

\section{Electron capture in the Thermal QRPA approach}

Due to the high temperature in the interior of massive stars, there is a finite probability
of occupation  of nuclear excited states in the stellar environment. We account for this by defining a thermal-averaged
cross section for capture of an electron with energy $\varepsilon_e$ on a particular nucleus
\begin{equation}\label{CrSect}
  \sigma(\varepsilon_e,T) = \sum_{if}p_i(T)\sigma_{if}(\varepsilon_e).
\end{equation}
Here, $p_i(T)$ is the  Boltzmann population factor for a parent state $i$ at temperature $T$, and $\sigma_{if}(\varepsilon_e)$ is
the cross section for capture of an electron from the  state $i$  to a state $f$ in the daughter nucleus. Then, the stellar
electron capture rate $\lambda(T)$  at finite temperature is obtained by folding  the thermal-averaged
cross section $\sigma(\varepsilon_e,T)$ with the distribution of
electrons,
\begin{equation}\label{ECrate}
  \lambda(T) = \frac{1}{\pi^2\hbar^3}\int^\infty_{m_ec^2}\sigma(\varepsilon_e,T)p_e^2f(\varepsilon_e)d\varepsilon_e,
\end{equation}
where $p_e=(\varepsilon_e^2-m_e^2c^4)^{1/2}/c$ is the momentum of the incoming electron. Under  conditions encountered
in the collapsing core of a supernova, electrons obey a Fermi-Dirac distribution $f(\varepsilon_e)$ with temperature $T$
and electron chemical potential $\mu_e$ which depend on the baryon density $\rho$ and the electron-to-baryon ratio $Y_e$.

The nuclei of interest in this study are expected to contribute to the stellar electron capture rates for temperatures $T\approx 0.5-1.5$~MeV.
At such high temperatures an explicit state-by-state evaluation of the sums in Eq.~\eqref{CrSect} is impossible with current nuclear models.
As was shown in~\cite{Dzhioev_PRC100}, within a statistical description, the thermal-averaged cross section can be expressed
through the temperature-dependent spectral functions  for the various momentum-dependent multipole operators
derived in~\cite{OConnell_PRC6,Walecka_1975}.

Although our calculations also consider forbidden transitions, we will give special emphasis to the
GT$_+$ contribution for the nuclear structure reasons outlined above.
Neglecting momentum transfer, the thermal-averaged cross section for
GT operators reduces to
\begin{align}\label{CrSect_GT}
  \sigma_\mathrm{GT}(\varepsilon_e,  T) =&  \frac{ G^2_\mathrm{F}
                                           V_\mathrm{ud}^2}{2\pi\hbar^4
                                           c^3 v_e} F(Z,\varepsilon_e)
  \notag \\
  &\times\int^{E_e}_{-\infty}(\varepsilon_e-E)^2 S_\mathrm{GT}(E,T) dE,
\end{align}
where $S_\mathrm{GT}(E,T)$ is the temperature-dependent strength function for the Gamow-Teller operator
\begin{equation}\label{SGT1}
   S_\mathrm{GT}(E,T) = \sum_{if} p_i(T)\frac{\bigl|\langle f\|g_A\pmb\sigma t_+\| i\rangle\bigr|^2}{2J_i+1}\delta(E- E_{if}).
\end{equation}
In the above equations $G_F$ is the  weak-interaction coupling constant, $g_A = -1.27$ is the axial-vector coupling constant,
and $V_{ud}$ is the up-down element in the Cabibbo-Kobayashi-Maskawa  quark mixing matrix.
The Fermi function $F(Z,\varepsilon_e)$ corrects the cross section for the distortion of the electron wave function by the
Coulomb field of the nucleus~\cite{Langanke_NPA673} and $v_e$ is the
electron velocity.
The transition energy between initial and final states is given by
$E_{if} =  Q + E_f-E_i$, where $E_{i,f}$ are the excitation energies  of the  parent and daughter nuclei, and $Q=M_f-M_i$ is the ground-state reaction threshold.
 At $T\ne 0$, due to transitions from thermally excited states, the strength function $S_\mathrm{GT}(E,T)$ is defined for
 both $E>Q$ and $E<Q$ domains.

To compute the temperature-dependent spectral functions we apply the TQRPA
framework~\cite{Dzhioev_IJMPhE18,Dzhioev_PhAN72,Dzhioev_PRC81,Dzhioev_PhAN79,Dzhioev_PRC100}, which  is a technique
based on the proton-neutron QRPA extended to finite temperature by the thermofield dynamics
formalism (TFD)~\cite{Takahashi_IJMPB10,Umezawa1982}.  The TFD doubles the degrees of freedom of the quantum system by introducing
a fictitious tilde Hamiltonian $\widetilde H$ and uses an extended Hilbert space of the direct product of the Hilbert spaces of
the physical and fictitious systems.
 The central concept in TFD is the thermal vacuum  $|0(T)\rangle$, a pure state in the extended Hilbert space,
 which corresponds to the thermal equilibrium, a mixed state in the original Hilbert space of the system.
 The time-translation operator in the extended Hilbert space is a so-called thermal Hamiltonian $\mathcal{H}=H-\widetilde H$.
 The temperature-dependent strength function is expressed by the transition matrix elements from  the thermal vacuum  to
 eigenstates  of the thermal Hamiltonian ($\mathcal{H}|Q_i\rangle = \omega_i |Q_i\rangle$):
  \begin{equation}\label{str_funct3}
    S_{A}(E,T)=
     \sum_{i} \bigl|\langle Q_i|\hat A|0(T)\rangle\bigr|^2\delta(E - \omega_i-\delta_{np}).
     \end{equation}
Here $\delta_{np}=\Delta\lambda_{np}+ \Delta M_{np}$, and $\Delta\lambda_{np}=\lambda_n-\lambda_p$ is the difference
between neutron and proton chemical potentials in the nucleus, and $\Delta M_{np}=1.293$~MeV is the neutron-proton mass splitting.
Note, that eigenvalues of the thermal Hamiltonian, $\omega_i$, take both positive and negative values. The latter contribute
to the strength function only at $T\ne 0$.

Within the TQRPA, the thermal Hamiltonian is diagonalized in terms of
thermal phonon operators, which are constructed as a linear
superposition of the creation and annihilation operators for
proton-neutron thermal quasiparticle pairs:
$\beta^\dag_p\beta^\dag_n$, $\beta^\dag_p\widetilde\beta^\dag_n$,
$\widetilde\beta^\dag_p\beta^\dag_n$,
$\widetilde\beta^\dag_p\widetilde\beta^\dag_n$, and their Hermitian
conjugates. Correspondingly thermal quasiparticles are connected with
Bogoliubov quasiparticles by the so-called thermal Bogoliubov
transformation which mixes nontilde and tilde operators.  It can be
shown~\cite{Dzhioev_PRC81} that the creation of a negative-energy
thermal tilde quasiparticle corresponds to annihilation of a thermally
excited Bogoliubov quasiparticles.  Because of single-particle
transitions involving annihilation of thermally excited Bogoliubov
quasiparticles, the phonon spectrum at finite temperature contains
states at negative- and low-energies which do not exist at zero
temperature and which correspond to thermally unblocked transitions of
excited nuclear states.  In the zero-temperature limit, the thermal
phonons reduce to the QRPA ones constructed of Bogoliubov
quasiparticle pairs $\alpha^\dag_p\alpha^\dag_n$ and
$\alpha_p\alpha_n$.

To analyze unblocking effects for EC in
neutron-rich nuclei with $N=50$, we perform Skyrme-TQRPA calculations
in $^{78}$Ni, $^{82}$Ge, $^{86}$Kr, and $^{88}$Sr.  To explore the
possible variations among parametrizations we have chosen two
different Skyrme parametrizations with sufficiently different
properties, SkM*~\cite{Bartel_NPA386} and
SkO$^\prime$~\cite{Reinhard_PRC60}. We solve Skyrme-Hartree-Fock
equations assuming spherical symmetry and neglecting thermal effects
on the mean-field. To take into account pairing correlations between like particles  we employ
the BCS pairing interaction. The pairing strength parameters are fixed
to reproduce the odd-even mass difference. Due to magic neutron number
$N=50$, there is no neutron pairing. Moreover, there is no proton
pairing in $^{78}$Ni and $^{88}$Sr, while proton pairing gaps in
$^{82}$Ge and $^{86}$Kr at $T=0$ are $\Delta_p=1.22$ and 1.28~MeV,
respectively. For very neutron-rich nuclei considered in the present work the difference between the neutron and proton chemical potentials is large enough to disregard isoscalar proton-neutron pairing in the mean field.

At $T\ne 0$ the pairing gap and the chemical potentials $\lambda_{n,p}$ are found from finite-temperature BCS
equations~\cite{Dzhioev_PRC81}.
The numerical solution of these equations yields vanishing pairing correlations
above a critical temperature $T_\mathrm{cr}\approx0.5\,\Delta$~\cite{Goodman_NPA352,Civitarese_NPA404}.
In the TQRPA calculations we use the Landau-Migdal force as the residual particle-hole interaction with the parameters derived from the Skyrme interaction~\cite{Giai_PLB379,Giai_PRC57}. We neglect the proton-neutron pairing in the residual interaction.   It is well known that the low-lying GT strength responsible for beta decay~\cite{Engel_PRC60} and double-beta decay~\cite{Engel_PRC37}  is sensitive to the isoscalar proton-neutron pairing interaction. However, the nuclei we consider dominate the nuclear composition of the collapsing core at high enough temperatures ($T>T_\mathrm{cr}$),  when particle-particle correlations  vanish and the residual proton-neutron pairing interaction does not affect the strength distributions.

\section{Results and discussion}\label{results}

\begin{figure}[t]
\includegraphics[width=\columnwidth]{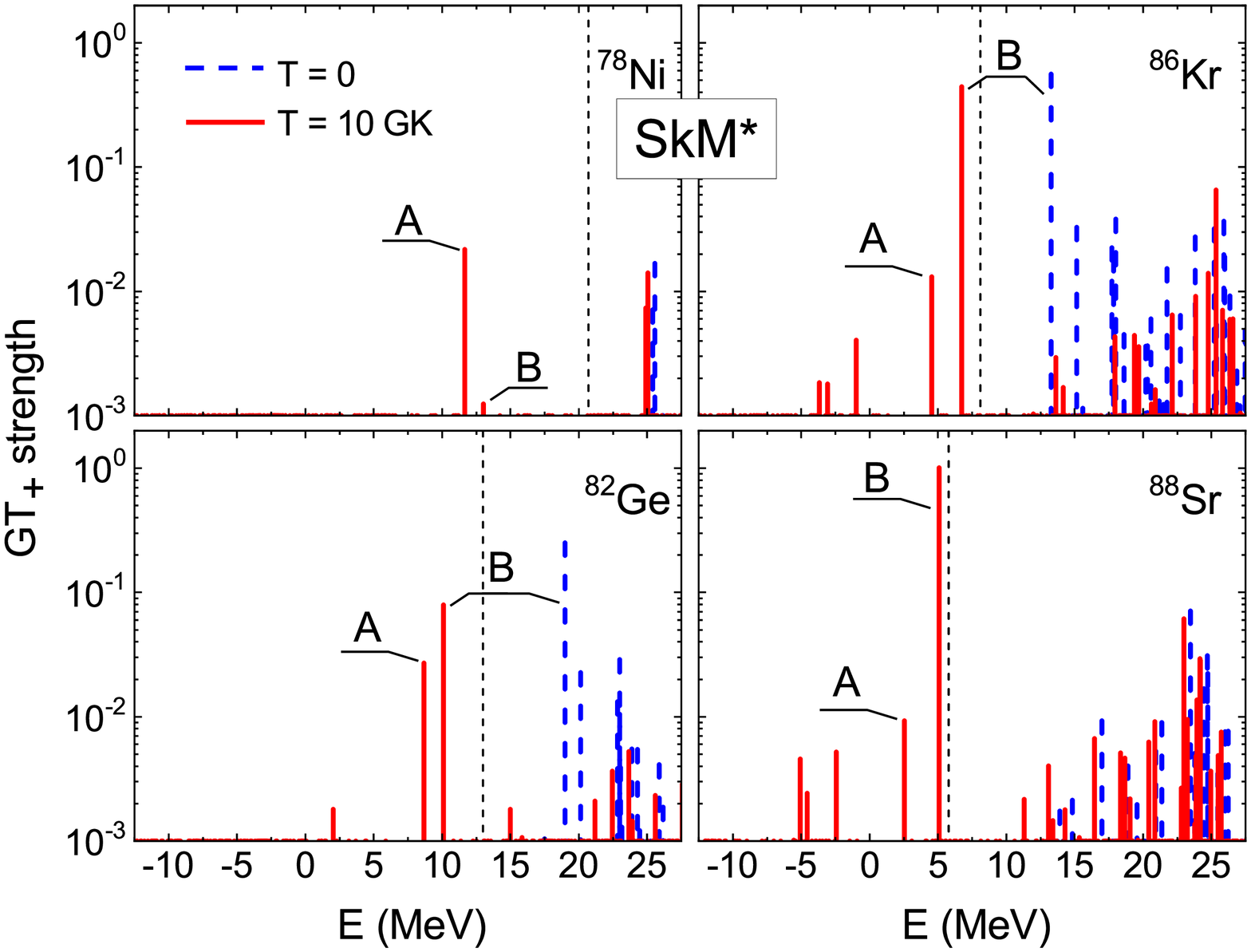}\vspace{3mm}
\includegraphics[width=\columnwidth]{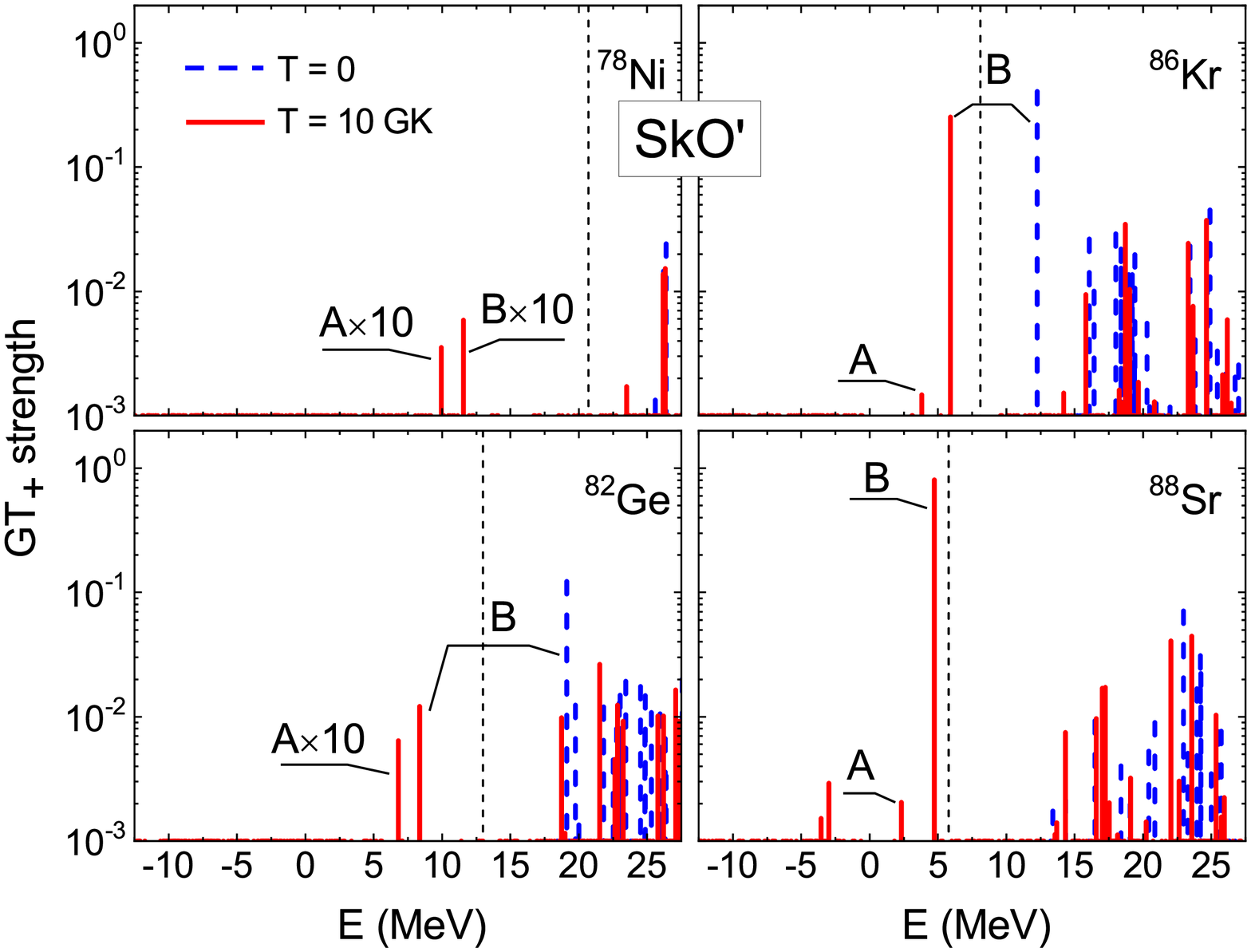}
\caption{(Color online) Strength distributions of GT$_+$ transitions in $^{78}$Ni, $^{82}$Ge, $^{86}$Kr, and $^{88}$Sr at $T=0$ and $T=10$~GK (0.86~MeV).
The dashed vertical lines indicate the ground state thresholds $Q=M_f-M_i$: $Q(^{78}\mathrm{Ni})=20.7~\mathrm{MeV}$, $Q(^{82}\mathrm{Ge})=13.0~\mathrm{MeV}$,
$Q(^{86}\mathrm{Kr})=8.1~\mathrm{MeV}$, $Q(^{88}\mathrm{Sr})=5.8~\mathrm{MeV}$~\cite{Moller_ADNDT109}. $A$ and $B$ label specific GT$_+$ transitions: $A\equiv f^p_{7/2}\to f^n_{5/2}$,
$B\equiv g^p_{9/2}\to g^n_{7/2}$. The notations $A\times 10$ and $B\times 10$ mean that the respective peaks are scaled by a factor of $10$ for demonstration purposes.}
\label{figure1}
\end{figure}

In this section we report about TQRPA calculations for neutron-rich
$N=50$ nuclei performed at temperatures $T=0$ and at $T=10$~GK (0.86~MeV), which is
a typical temperature at which these nuclei are abundant in collapsing
supernova cores. Our main attention is laid upon the GT$_+$ response and its
unblocking at finite temperatures, which is particularly important for the
electron capture rates on these nuclei under supernova conditions.
Finally we supplement the GT part of the electron capture rate by
the contributions of the other multipoles.

Figure~\ref{figure1} shows our GT$_+$ response at $T=0$ calculated
with the Skyrme interaction SKM$^*$. The results obtained
with  SkO$^\prime$ are qualitatively similar. As GT$_+$ transitions are
completely blocked for nuclei with $Z<40$ and $N>50$, the calculated
transitions are due to nuclear correlations induced by the Skyrme and
the pairing forces. In our calculations the main unblocking mechanism
in the ground states is due to the excitations of protons into the
$g_{9/2}$ orbital, enabling GT$_+$ transitions into the $g_{7/2}$
neutron orbitals. However, these transitions reside at relatively
modest excitation energies, resulting in the fact that there is no
GT strength at low excitation energies for the ground state distribution.
This observation is in agreement with the measured GT$_+$ strengths for
$^{86}$Kr and $^{88}$Sr, which both find no strength at low energies
$E < 7$ MeV~\cite{Zamora_PRC100,Titus_arXiv1908}.
We note that the correlations across the $Z=40$ and $N=50$ shell
gaps open up other possible GT$_+$ transitions, for example between
$f_{7/2}$ proton and $f_{5/2}$ neutron orbitals. But these transitions
are small compared with $g^p_{9/2} \rightarrow g^n_{7/2}$ and reside
at slightly higher excitation energies. We note that the
$g^p_{9/2} \rightarrow g^n_{9/2}$ transition, which corresponds
to rather small excitation energies and would be important for
electron capture, requires the excitations of protons and neutrons
across the shell gaps and hence is doubly suppressed in our model.
Hence we do not find any relevant strength for this transition.

The ground state ($T=0$) GT$_+$ strength for the $g^p_{9/2} \rightarrow g^n_{7/2}$ transition
(and for the other transitions) is noticeably larger for $^{82}$Ge and
$^{86}$Kr than for $^{78}$Ni and $^{88}$Sr. This is related to the
fact that proton pairing is absent in the latter two nuclei and
non-vanishing GT$_+$ strength in $^{78}$Ni and $^{88}$Sr appears only
at relatively high energies due to the admixture of $2\hbar\omega$
correlations.  For $^{82}$Ge and $^{86}$Kr configuration mixing is
induced by the pairing interaction which mixes 0p0h and 2p2h
configurations.  For $^{82}$Ge and $^{86}$Kr, our BCS-SkM*
calculations predict the occupation numbers $\langle n\rangle=0.2$
and~0.4 protons in the $g^p_{9/2}$ orbital, respectively, thus making
possible the $g^p_{9/2}\to g^n_{7/2}$ transition. The larger
occupation number for $^{86}$Kr reflects itself in the larger strength
of the $g^p_{9/2}\to g^n_{7/2}$ transition, as seen in
Fig. \ref{figure1}.  At $T=0$, this transition corresponds to the
excitation of the $\alpha^\dag_{g^p_{9/2}}\alpha^\dag_{g^n_{7/2}}$
configuration above the ground state.  The respective transition
strength is proportional to the BCS amplitude $v^2_{g^p_{9/2}}$, while the transition energy is determined the sum
of quasiparticle energies
$\varepsilon_{g^n_{7/2}}+\varepsilon_{g^p_{9/2}}+\delta_{np}$.

At finite temperature the thermally averaged GT$_+$ strength
is arising from an ensemble of excited nuclear states, where the
centroid and the width of the ensemble increases with growing temperature.
At the temperatures of about $T=1 $~MeV, at which the neutron-rich
$N=50$ nuclei are expected to contribute to the stellar electron capture
rate, the structure of these states involve a larger occupation
of particle orbitals  above those occupied in the ground state, leaving at the
time holes in orbitals occupied at $T=0$. This mechanism thermally unblocks
GT$_+$ transitions; for example, the
$g^p_{9/2}\to g^n_{7/2}$ and $f^p_{7/2} \to f^n_{5/2}$ transitions, which
in our TQRPA formalism are accompanied by
annihilation of thermally excited particle and hole states.
The resulting configurations above the thermal vacuum
are  $\widetilde\beta^\dag_{g^p_{9/2}}\beta^\dag_{g^n_{7/2}}$,
$\beta^\dag_{f^p_{7/2}}\widetilde\beta^\dag_{f^n_{5/2}}$,
while the transition strengths  are proportional to
$y^2_{g^p_{9/2}}(1-y^2_{g^n_{7/2}})$, $(1-y^2_{f^p_{7/2}})y^2_{f^n_{5/2}}$. Here $y^2_j=[1+\exp(-\varepsilon_j/T)]^{-1}$ is the thermal
occupation factor for a single-particle state $j$.
Within the TFD approach this factor stems from the thermal
Bogoliubov transformation.
We stress that the energies of the thermally unblocked transitions are
$\varepsilon_{g^n_{7/2}}-\varepsilon_{g^p_{9/2}}+\delta_{np}$, $\varepsilon_{f^p_{7/2}}-\varepsilon_{f^n_{5/2}}+\delta_{np}$,
which differs from the respective transition energies at $T=0$ by the minus
signs which correspond  to annihilation of thermally excited states.
Our calculated finite-temperature GT$_+$ strengths are shown in Fig.~\ref{figure1}
together with the ground state distributions for comparison.
For all considered nuclei, thermally unblocked GT$_+$ transitions are
located  below the ground-state threshold~$Q$, including the
strength due to the dominant
$g^p_{9/2}\to g^n_{7/2}$ and $f^p_{7/2} \to f^n_{5/2}$ transitions.
This appearance of low-lying strength will have important consequences
for the stellar electron capture rates at finite temperatures.
Moreover, for $^{86}$Kr and $^{88}$Sr we observe some
thermally unblocked GT$_+$ strengths even at negative energies,
which correspond to transitions from thermally excited states in the parent nucleus that are at higher energies than the final
states in the daughter nucleus. Due to negative energy transitions, nuclear excitation energy is transferred to the outgoing neutrinos.

Except for $^{78}$Ni, the largest GT$_+$ strength resides in peaks
which correspond mainly to
$g^p_{9/2}\to g^n_{7/2}$ transitions. Its strength increases from $^{82}$Ge
to $^{88}$Sr, related to the number of protons in the $pf$ shell available
for thermal excitations into the $g_{9/2}$ orbital.
We also note that pairing correlations vanish with temperature
and are already strongly diminished at $T=10$~GK. As a consequence
the noticeable peaks seen in the ground-state GT$_+$ distributions for $^{82}$Ge and $^{86}$Kr
are suppressed in the TQRPA calculation at $T=10$~GK. Referring to Fig.~\ref{figure1}, at $T=10$~GK the total GT$_+$ strength in $^{82}$Ge and $^{86}$Kr appears to be somewhat smaller than that at $T=0$. The origin of non-monotonic temperature dependence of the  total GT$_+$ strength in neutron-rich nuclei is discussed in Ref.~\cite{Dzhioev_PRC81}. In particular  it is shown that the total strength  reaches a minimum value in the vicinity of the critical temperature $T\approx 0.5\Delta$, i.e., when pairing correlations vanish, but thermal effects are not yet sufficiently strong to occupy the $1g_{9/2}$ proton orbit and unblock the $1f_{5/2}$ neutron orbit. Note however, that the Ikeda sum rule is fulfilled within the TQRPA~\cite{Dzhioev_PhAN72}.

\begin{figure}[t]
\includegraphics[width=\columnwidth]{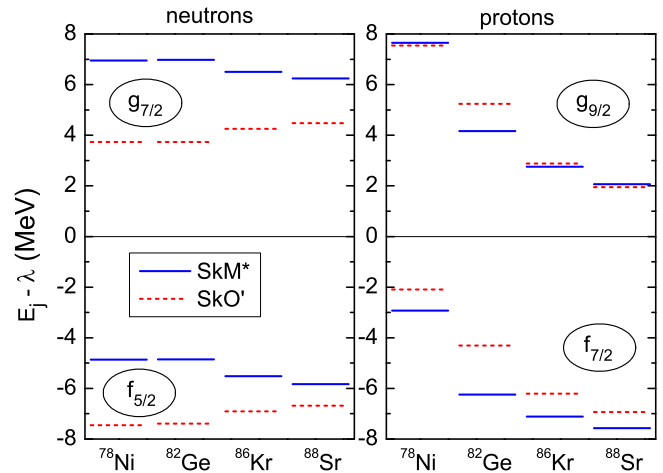}
\caption{ (Color online) Neutron ($f_{5/2}$, $g_{7/2}$) and proton ($f_{7/2}$, $g_{9/2}$)  single-particle energies $E_j$ relative to the chemical potentials $\lambda_{n,p}$ for $T=10$~GK. The  respective quasiparticle energies are given by $\varepsilon_j=|E_j-\lambda|$.}\label{figure4}
\end{figure}

Figure~\ref{figure1} reveals that the strength of thermally unblocked $f^p_{7/2}\to f^n_{5/2}$
transitions depends rather noticeably on the Skyrme parametrization, but varies only moderately among
the different nuclei.  In contrast, the peak related to the thermally unblocked $g^p_{9/2}\to g^n_{7/2}$ transition
increases by almost three orders of magnitude between $^{78}$Ni and $^{88}$Sr, but
it is rather insensitive to the choice of the Skyrme parametrization.
To gain insight into this observation we plot in Fig.~\ref{figure4} the single-particle energies $E_j$
for the particle ($g^p_{9/2}$, $g^n_{7/2}$) and hole ($f^p_{7/2}$, $f^n_{5/2}$) orbitals
relative to the chemical potentials $\lambda_{n,p}$ at $T=10$~GK, as calculated for the two Skyrme interactions.
Note that for $T>T_\mathrm{cr}$ the quasiparticle energy is given  by $\varepsilon_j =  |E_j-\lambda|$.
Hence $|E_j-\lambda|$ determines the occupation probabilities $y^2_j$ of the hole (particle) orbitals.
For $T=10$~GK, the proton orbital $f^p_{7/2}$ remains almost occupied (i.e., $1-y^2_{f^p_{7/2}}\approx 1$)
and  the strength of the $f^p_{7/2}\to f^n_{5/2}$ transition mainly depends on the number of thermally excited vacancies
in the neutron orbital $f^n_{5/2}$.
Figure~\ref{figure4} shows that the Skyrme interaction SkO$^\prime$  predicts a larger quasiparticle energy  $\varepsilon_{f^n_{5/2}}=|E_{f^n_{5/2}}-\lambda_n|$
than the SkM$^*$ interaction. However, $E_{f^n_{5/2}}-\lambda_n$ does not change significantly with increasing proton number.
As a consequence, the occupation factors $y^2_{f^n_{5/2}}$ obtained with the SkM$^*$ force are  larger
than those calculated with  the SkO$^\prime$ Skyrme interaction, but they do not vary much between the different
nuclei.
For the $g^p_{9/2}\to g^n_{7/2}$ transition, the thermally unblocked strength is mainly determined by the occupation
of the $g^p_{9/2}$ orbital. Referring to Fig.~\ref{figure4}, the quasiparticle energy $\varepsilon_{g^p_{9/2}}=E_{g^p_{9/2}}-\lambda_p$ reduces by a factor of four between $^{78}$Ni and $^{88}$Sr.
This reduction increases the occupation factor $y^2_{g^p_{9/2}}$, and hence the transition strength,  by almost three orders of magnitude.
Both Skyrme parametrizations predict rather
similar values for $E_{g^p_{9/2}}-\lambda_p$.

\begin{figure}[t]
\includegraphics[width=\columnwidth]{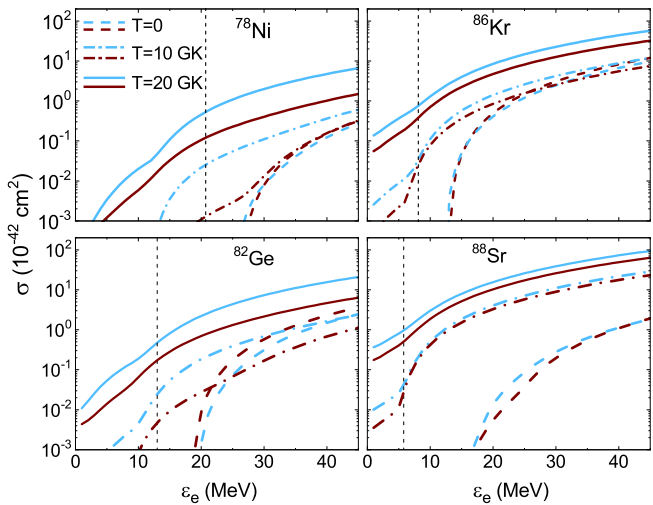}
\caption{(Color online) Electron capture cross sections for $^{78}$Ni, $^{82}$Ge,
  $^{86}$Kr, and $^{88}$Sr at $T=0,~10,~20$~GK.  The blue (light gray) and brown (dark gray)
  curves represent results obtained with the SkM$^*$ and SkO$^\prime$
  interactions, respectively.  Like in Fig.~\ref{figure1}, the dashed vertical lines
  indicate the ground state thresholds $Q=M_f-M_i$.}\label{figure2}
\end{figure}

The thermal unblocking of GT$_+$ transitions in the neutron-rich $N=50$ nuclei
reflects itself strongly in the GT contributions to the electron capture cross sections.
In Fig.~\ref{figure2} we plot thermal-averaged electron capture cross sections at $T=10$~GK and at $T=20$~GK
as a function of the incident electron energy $\varepsilon_e$ in comparison with the ground-state ($T=0$) results.
In our calculations of the cross sections (and rates) we account for the quenching of the GT strength by the reduction of the
axial-vector coupling constant $g_A$. In the context of shell model calculations that really test individual GT transitions the quenching factor has been determined to  be 0.74 for $pf$-shell nuclei~\cite{Caurier&Zuker_PRC50,Langanke&Dean_PRC52}.
With the same quenching factor, as shown in Ref.~\cite{Dzhioev_PRC81},  the experimental GT$_+$ strength in  $pf$-shell nuclei can be reproduced by the QRPA calculations.
For all nuclei and for both interactions we observe a similar overall  evolution of the cross sections with increasing temperature,
reflecting the
temperature dependence of the GT$_+$ strength distributions as discussed above.
For all nuclei, the ground-state GT$_+$ distributions have no strength at low energies, representing a threshold for
electron captures. Hence the $T=0$ cross sections vanish at small electron energies. Once the
capture threshold is overcome, the cross sections increase with electron energies, where the energy dependence is mainly
dictated by phase space.
With increasing temperatures, GT$_+$ strength
is shifted towards lower energies (see Fig. \ref{figure1}), reducing the gap which has to be overcome or even removing
it completely if strength is shifted to negative energies due to downscattering transitions (see discussion above).
For the nuclei studied here, a gap exists in $^{78}$Ni, even at $T= 20$~GK.
For the other nuclei, the gap has vanished at this temperature as the reaction threshold completely disappears
due to the contribution of GT$_+$ strength at negative energies.

The detailed energy dependence of the GT$_+$ strength is decisive at low electron energies, but becomes less relevant with
increasing energy, where, however, the total strength matters~\cite{Juodagalvis_NPA848}. In the two cases with proton
pairing ($^{82}$Ge, $^{86}$Kr) there is noticeable strength at moderate energies in our model calculations for $T=0$
and the total strength is somewhat larger for the ground state distribution than at $T=10$~GK (see our discussion above). As a consequence the capture cross sections
at high electron energies is slightly larger than at $T=10$~GK. This phenomenon does not occur for $^{78}$Ni and $^{88}$Sr,
where no proton pairing effects exist to create cross-shell correlations and hence the total ground state strength
is small. Relatedly, the increase of the cross sections with temperature is largest for these two nuclei.
We also note that the differences in cross sections for the various nuclei become smaller with increasing electron energies,
as had already been observed and explained in Refs.~\cite{Juodagalvis_NPA848,Dzhioev_PRC81}. At the finite temperatures
we have studied here, the cross sections at high energies increase with number of protons as the promotion of protons in the $g_{9/2}$
orbital is the main unblocking mechanism.
Comparing the results obtained with the SkM$^*$ and SkO$^\prime$ forces we conclude that the most essential differences
in the cross sections exist in more neutron-rich nuclei at low temperatures
and they reflect the differences in the GT$_+$ strength distributions discussed above.

\begin{figure}[t]
  \includegraphics[width=\columnwidth]{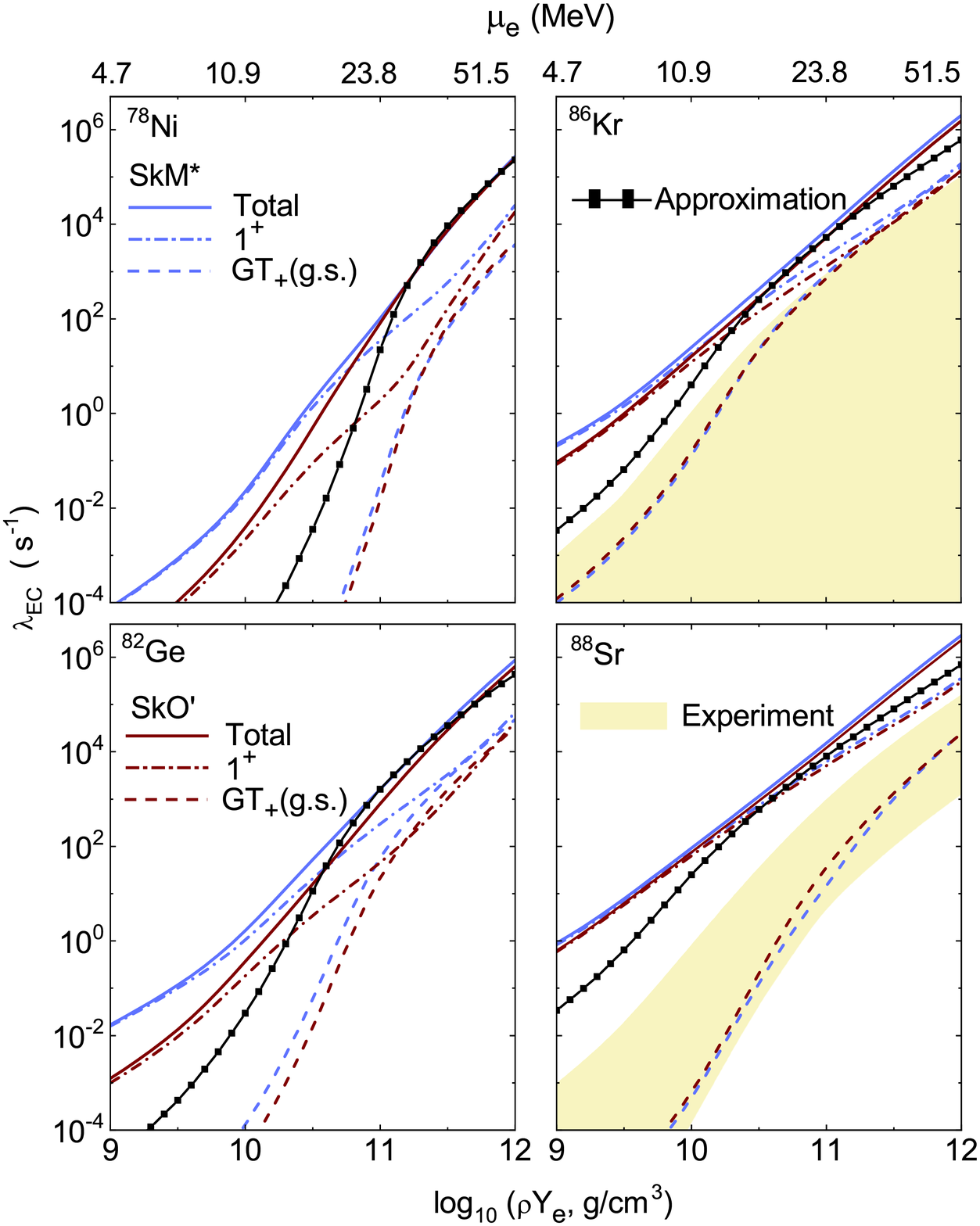}
\caption{(Color online) Electron capture rates calculated at $T=10$~GK (0.86~MeV) as
  a function of density. The upper axis indicates the corresponding
  electron chemical potentials~$\mu_e$.
  The blue (light gray) and brown (dark gray) lines are based on the Skyrme-TQRPA calculations with the SkM$^*$
  and SkO$^\prime$ interactions, respectively.
  The total rates (full lines) include the
  contribution of allowed ($0^+,~1^+$) and first-forbidden
  ($0^-,~1^-, 2^-$) transitions. The dashed-dotted lines (labelled $1^+$) represent
  the unblocked GT$_+$ contributions to the rates. The rates, indicated by dashed lines, have been calculated from the ground-state GT$_+$ distribution.
  The shaded bands represent the results based
  on the experimental GT$_+$ data
  \cite{Titus_arXiv1908,Zamora_PRC100}. The labeled lines represent
  the rates calculated according to the
  parametrization~\eqref{EC_param}.  }\label{figure3}
\end{figure}

Figure~\ref{figure3} shows electron capture  rates for  $^{78}$Ni, $^{82}$Ge, $^{86}$Kr, and $^{88}$Sr at temperature $10$~GK and for
densities of relevance for the collapse phase of core-collapse supernova.\footnote{For example,
the core temperatures are in the range $10-14$~GK for densities $\rho Y_e$ between
$2.4\times10^{10}$ g~cm$^{-3}$ and $2.3 \times 10^{11}$ g~cm$^{-3}$ for a $15 M_\odot$ star due to Table 1 in Ref.~\cite{Juodagalvis_NPA848}}.
Our rates are obtained by integrating the thermal-averaged cross sections following  Eq.~\eqref{ECrate}.
In our rate calculations we have considered
spectral functions for the allowed ($0^+$,~$1^+$) and first-forbidden ($0^-$, $1^-$, $2^-$)
momentum-dependent multipole operators derived in~\cite{OConnell_PRC6, Walecka_1975}. To demonstrate the relevance of thermal unblocking
and of multipoles other than Gamow-Teller, the figure also exhibits EC rates calculated from the ground-state and finite-temperature
GT$_+$ distributions. For comparison we also show the capture rates for $^{86}$Kr and $^{88}$Sr  derived from
the experimental Gamow-Teller data~\cite{Zamora_PRC100,Titus_arXiv1908}. The shaded bands are due to the experimental
uncertainty in the GT$_+$ strength.

We note that the EC rates for $^{86}$Kr and $^{88}$Sr derived from the ground-state QRPA
calculations are consistent with the one derived from the experimental data. However,
the rates obtained at finite temperatures ($T=10$~GK) are significantly larger than those obtained from the ground state
distributions stressing the importance of thermal unblocking effects in our calculation. This holds at all densities,
but it is most pronounced at low densities where smaller electron energies have relatively more weight.
The contribution from the GT$_+$ strength dominates the rates at lower densities (again due to the smaller electron
energies involved). The contributions from forbidden multipoles becomes increasingly relevant with growing
densities. At densities in excess of a few $10^{10}$ g~cm$^{-3}$ they dominate the rates (see also Table~\ref{forb_contr}). As the forbidden transitions
are rather insensitive to the differences in single-particle energies obtained for the two
Skyrme parametrizations and to thermal effects~\cite{Dzhioev_PRC100}, the rates are also very similar for the two Skyrme interactions at the high densities.

\begin{table}
\caption{Relative contribution, $\lambda_\mathrm{EC}^{ff}/\lambda_\mathrm{EC}$, of first forbidden transitions to the electron capture rates at $T=10$~GK and selected densities $\rho Y_e$ (in g\,cm$^{-3}$). The results are obtained with the SkM* (SkO$^\prime$) Skyrme interaction. }
\label{forb_contr}
\begin{tabular}{ccccc}
\hline\noalign{\smallskip}
$\log_{10}(\rho Y_e)=$          &9 &     10      &   11        & 12         \\
\noalign{\smallskip}\hline\noalign{\smallskip}
$^{78}$Ni~~ &   0.04 (0.17)~ & 0.14 (0.44)~ & 0.66 (0.97)~ & 0.91 (0.93) \\
$^{82}$Ge~~ &   0.07 (0.20)~ & 0.37 (0.51)~ & 0.82 (0.94)~ & 0.92 (0.93) \\
$^{86}$Kr~~ &   0.09 (0.09)~  & 0.27 (0.25)~ & 0.71 (0.75)~ & 0.90 (0.91) \\
$^{88}$Sr~~ &   0.08 (0.06)~  & 0.18 (0.16)~ & 0.60 (0.59)~ & 0.87 (0.86) \\
\noalign{\smallskip}\hline
\end{tabular}
\end{table}

Thus, the main result from  Fig. \ref{figure3} is that the derivation of stellar capture rates for
the neutron-rich $N=50$ nuclei on the basis of the ground state GT$_+$ distributions is unjustified.
During the stellar collapse, such nuclei are quite abundant at densities of order $10^{11}$ g~cm$^{-3}$. As can be read
off from Fig. \ref{figure3}, thermal unblocking and the contributions from forbidden multipoles enhance the capture
rates by an order of magnitude or more.

In Refs.~\cite{Langanke_PRC63,Langanke_PRL90,Juodagalvis_NPA848} a hybrid model has been introduced and
used to derive stellar electron capture rates at densities in access of $10^{10}$ g~cm$^{-3}$. In the first step
the Shell Model Monte Carlo (SMMC) method~\cite{Johnson_PRL69} is used to calculate partial occupation numbers
at finite temperatures taking multi-nucleon correlations into account by a residual interaction acting
in large model spaces. In the second step, these partial occupation numbers serve as input into the calculation of stellar
capture rates using an RPA approach. The studies within the hybrid model show that multi-nucleon correlations induced by the
residual interaction and by thermal excitation are strong enough to overcome the shell gaps at finite temperature
and to unblock the GT$_+$ strength. Figure~1 of Ref. \cite{Langanke_PRL90} shows electron capture rates
derived within the hybrid model for selected nuclei, including $^{89}$Br ($Z=35$ and $N=54$). The figure clearly indicates
that within the hybrid model shell gaps, including the one at $N=50$,  are overcome at astrophysical conditions (temperatures)
present during collapse at densities in access of $10^{11}$ g~cm$^{-3}$ (corresponding to electron chemical potentials
$\mu_e > 15$ MeV~\cite{Juodagalvis_NPA848}).

Ref.~\cite{Dzhioev_PRC81} has performed detailed comparisons between
electron capture rates calculated for the germanium isotopes
$^{76,78,80}$Ge within the hybrid model and the present TQRPA
approach. As can be seen in Fig. 9 of Ref. \cite{Dzhioev_PRC81} there
are noticeable differences between the two approaches for the
conditions at low densities where details of the GT$_+$ strength
distribution, induced by the different treatment of cross-shell
correlations, still matter. For the reasons explained above and in
Ref.~\cite{Juodagalvis_NPA848}, these differences decrease with
increasing density and temperature.  We also note that the capture
rate decreases with increasing neutron number in the germanium
isotopes. This is mainly due to the increasing $Q$ value, which has to
be overcome by the capture process, and the increasing number of
neutrons partially blocking transitions into the neutron $g_{9/2}$
orbital. If we take the rates obtained at $T=10$~GK (0.86~MeV) from
Fig.~9 of Ref.~\cite{Dzhioev_PRC81}, our present capture rate for
$^{82}$Ge obtained for $\rho Y_e$ = $10^{11}$ g~cm$^{-3}$
($\lambda_{ec} \approx 10^3~\mathrm{s}^{-1}$) agrees nicely with the
trend of the rates for $^{76,78,80}$Ge calculated with the TQRPA and
the hybrid model.

In Ref. \cite{Langanke_PRL90} a rather simple parametrization for the
capture rate has been derived by fit to individual electron capture
rates available at that time (about 200 nuclei in the mass range
$A = 45-110$. The purpose of the study presented in
\cite{Langanke_PRL90} was to demonstrate that the $N=40$ shell gap
does not block the GT$_+$ strength in neutron-rich nuclei and that
electron capture during the later phase of the collapse proceeds on
nuclei and not on free protons, as had been hypothized earlier in the
investigation of Ref.  \cite{Langanke_PRL90}. For this goal, the pool
of about 200 nuclei, for which individual rates at their relevant
astrophysical conditions have been evaluated and used in the supernova
simulation, was sufficient.  As the fit formula was rendered too
simple, further individual capture rates have been derived in
Ref.~\cite{Juodagalvis_NPA848} based on a hierarchical structure
approach suitable to the large set of nuclei considered. Each nucleus
is described by a model which is thought to be accurate enough at the
astrophysical conditions at which the nuclei contribute to the overall
capture rate. The respective rate table derived by Juodagalvis and
collaborators is being used in modern supernova simulation codes
(see, e.g., refs~\cite{Janka2012,Kotake2018}).

Unfortunately the rate table of Ref. \cite{Juodagalvis_NPA848}
does not exist for individual nuclei, but rather for an ensemble of nuclei
distributed in nuclear statistical equilibrium. Hence authors have
recently returned to the fit formula of Ref.~\cite{Langanke_PRL90}
to explore sensitivity
of supernova simulations to certain input parameters
\cite{Raduta_PRC93,Raduta_PRC95, arXiv_Pascal}. As this formula has been derived
to a pool of nuclei which does not include neutron-rich nuclei
at the $N=50$ shell gap, a comparison to the present TQRPA rates
for such nuclei is quite instructive. The simple formula is based on
the single-state approximation and reads~\cite{Langanke_PRL90}
\begin{equation}\label{EC_param}
  \lambda = \frac{B \ln 2}{K}\Bigl(\frac{T}{m_e c^2}\Bigr)^5\bigl[ F_4(\eta) - 2\chi  F_3(\eta)  + \chi^2 F_2(\eta)\bigr].
\end{equation}
Here $K=6146$~s, $F_k(\eta)$ are the Fermi integrals of rank $k$ and
degeneracy $\eta$, $\chi = -(Q+\Delta E)/T$~
\footnote{Note that in our definition $Q=M_f - M_i$, while in~Ref.~\cite{Langanke_PRL90}
the $Q$ value is defined with opposite sign.},
and $\eta =\chi + \mu_e/T$.
The fit parameters $B=4.6$ and $\Delta E=E_f-E_i =2.5$~MeV represent
effective values for the transition strength
and the energy difference between the final and initial excited states,
respectively. The pool of nuclei to which the fit has been performed
included $pf$ shell nuclei and some heavier nuclei with $A<100$.
For the $pf$ shell nuclei, which dominate the captures at
lower densities $\rho < 10^{10}$ g~cm$^{-3}$, the rates only include
Gamow-Teller transitions taken from diagonalization shell model
calculations~\cite{Langanke_NPA673}, while the heavier nuclei,
which are relevant at densities above $10^{10}$ g~cm$^{-3}$ also include
forbidden contributions.

In Fig. \ref{figure3} we compare our TQRPA results to the fit formula.
We observe that at the densities $\rho Y_e \approx 10^{11}$ g~cm$^{-3}$
where the neutron-rich $N=50$ nuclei are relevant, the fit
reproduces our TQRPA capture rates quite well. This again shows that at
these late-collapse conditions the capture rates are rather insensitive
to details of the nuclear response. This is not true at lower densities.
For example at $\rho Y_e =10^{10}$ g~cm$^{-3}$ the fit underestimates
the TQRPA capture rates by an order of magnitude. The insufficiency
of the fit under such conditions had already been discussed
before (see, e.g., \cite{Raduta_PRC93}). However, at these low densities
the fit formula should not be used because the rates are still sensitive
to details of the strength distributions, in particular to nuclei
with rather large $Q$ values like the neutron-rich $N=50$ nuclei. But
importantly, these nuclei are quite unabundant at these low-density
conditions and hence do not contribute to the overall capture rates.

\section{Conclusion}\label{conclusion}

We have studied the electron capture rates on neutron-rich $N=50$
nuclei at conditions of temperatures and densities relevant for
collapse supernovae. Our calculations have been motivated by the
suggestion that the $N=50$ shell gap could serve as an obstacle for
electron captures in supernovae \cite{Zamora_PRC100,Titus_arXiv1908}
blocking GT$_+$ transitions. In fact, experimental GT$_+$
distributions obtained for the $N=50$ nuclei $^{86}$Kr and $^{88}$Sr
do not show any strength at low energies. Our $T=0$ QRPA calculations,
performed for these two nuclei and $^{78}$Ni and $^{82}$Ge, reproduce
this observation, not showing strength at low energies $E <7$ MeV in
any of these nuclei, in agreement with Pauli blocking of the GT$_+$
strength for $N=50$ nuclei. However, our finite temperature TQRPA
calculations also show that this blocking is overcome at finite
temperatures due to thermal excitations, enabling transitions from
proton $f_{7/2}$ and $g_{9/2}$ orbitals into neutron $f_{5/2}$ and
$g_{7/2}$ orbitals, respectively. Both noticeably unblock the GT$_+$
strength at supernova conditions where these nuclei are abundant. Our
calculations also indicate that at the corresponding relatively high
density conditions forbidden transitions contribute significantly to
the capture rates.  The unblocking of the GT$_+$ strength at finite
temperatures and the sizable forbidden contributions imply that the
derivation of stellar electron capture rates for the neutron-rich
$N=50$ nuclei on the basis of the GT$_+$ ground state distribution, as
presented in \cite{Zamora_PRC100}, is inappropriate. Our results also
indicate that the neutron-rich $N=50$ nuclei do not act as obstacles
for electron captures in the later collapse phase.  Our results at the
relevant astrophysical conditions are in good agreement with those
obtained in the hybrid model proposed in \cite{Langanke_PRL90} which
is the basis of the electron capture rate tables
\cite{Juodagalvis_NPA848} presently in use in supernova simulations.

\begin{acknowledgments}
  KL and GMP are partly funded by the Deutsche Forschungsgemeinschaft
  (DFG, German Research Foundation) -- Project-ID 279384907 -- SFB
  1245.  Part of this work was done while the first author visited the
  GSI Helmholzzentrum f\"{u}r Schwerionenforschung.  He is grateful
  for the warm hospitality and the financial support.
\end{acknowledgments}

\bibliography{library}


\end{document}